\documentclass[prd,aps,twocolumn,nofootinbib,preprintnumbers,superscriptaddress,preprintnumbers,balancelastpage,longbibliography]{revtex4-2}
\pdfoutput=1
\usepackage{amsmath,mathtools,physics,xfrac}
\usepackage{graphicx}
\usepackage{afterpage}
\usepackage{float}
\usepackage{subfigure}
\usepackage{rotating}
\usepackage{multirow}
\usepackage{tabularx}
\usepackage{booktabs}
\usepackage{fancyhdr}
\usepackage[utf8]{inputenc}
\usepackage{theorem}
\usepackage{moreverb}
\usepackage{euscript}
\usepackage{psfrag}
\usepackage{slashed}
\usepackage{mathtools}
\usepackage{makecell}
\usepackage{adjustbox}
\usepackage{dcolumn}
\usepackage{bm}
\usepackage[dvipsnames]{xcolor}
\usepackage{graphics}
\usepackage{hyperref}
\usepackage{lettrine}
\usepackage[shortlabels]{enumitem}

\input Zallman.fd

\LettrineTextFont{\itshape}
\definecolor{nice}{rgb}{0.996, 0.59, 0.035}

\hypersetup{
     colorlinks   = true,
     citecolor    = nice,
     urlcolor     = nice,
     linkcolor    = nice
}

\begin{document}
\preprint{SLAC-PUB-17770}

\title{Dark Branches of Immortal Stars at the Galactic Center}

\author{Isabelle John}
\thanks{\href{mailto:isabelle.john@fysik.su.se}{isabelle.john@fysik.su.se}, \href{http://orcid.org/00000-0003-2550-7038}{ORCID: 0000-0003-2550-7038}}
\affiliation{Stockholm University and The Oskar Klein Centre for Cosmoparticle Physics,  Alba Nova, 10691 Stockholm, Sweden}
\author{Rebecca K. Leane}
\thanks{\href{mailto:rleane@slac.stanford.edu}{rleane@slac.stanford.edu}, \href{http://orcid.org/0000-0002-1287-8780}{ORCID: 0000-0002-1287-8780}}
\affiliation{Particle Theory Group, SLAC National Accelerator Laboratory, Stanford, CA 94035, USA}
\affiliation{Kavli Institute for Particle Astrophysics and Cosmology, Stanford University, Stanford, CA 94035, USA}\author{Tim Linden}
\thanks{\href{mailto:linden@fysik.su.se}{linden@fysik.su.se}, \href{http://orcid.org/0000-0001-9888-0971}{ORCID: 0000-0001-9888-0971}}
\affiliation{Stockholm University and The Oskar Klein Centre for Cosmoparticle Physics,  Alba Nova, 10691 Stockholm, Sweden}

\begin{abstract}
\noindent We show that stars in the inner parsec of the Milky Way can be significantly affected by dark matter annihilation, producing population-level effects that are visible in a Hertzsprung-Russell (HR) diagram. We establish the dark HR diagram, where stars lie on a new stable \emph{dark main sequence} with similar luminosities, but lower temperatures, than the standard main sequence. The dark matter density in these stars continuously replenishes, granting these stars immortality and solving multiple stellar anomalies. Upcoming telescopes could detect the dark main sequence, offering a new dark matter discovery avenue.
\end{abstract}

\maketitle

\lettrine{S}{tars evolve} along well-established tracks within which an equilibrium between the gravitational force and the pressure from nuclear fusion can be maintained. Their blackbody spectrum also directly links their luminosity, temperature and radius. These relationships are apparent on the Hertzprung-Russell (HR) diagram, which plots the observable stellar luminosities and temperatures, tracing various mass-dependent tracks stars travel across over their lifetime.

This standard picture of stellar evolution assumes that nuclear fusion is the only process capable of providing pressure support for the star. However, in the early Universe~\cite{Spolyar:2007qv, Iocco:2008xb,Freese:2008hb, Taoso:2008kw, Sivertsson:2010zm, Freese:2015mta,Ilie:2020iup, Ilie:2020nzp}, or in regions very close to the Galactic Center~\cite{1989ApJ...338...24S,Fairbairn:2007bn, Scott:2008ns,Lopes:2021jcy}, dark matter annihilation may provide a considerable additional energy source. In extreme cases, dark matter annihilation can cause stars to move backwards along the HR diagrams towards proto-stellar configurations (Hayashi track) that have extremely large radii and low temperatures compared to standard stars~\cite{Scott:2008ns}. In some cases, dark matter annihilation can cause stars to be disrupted entirely~\cite{John:2023knt}.

In this article we produce, for the first time, a population-level description of main sequence evolution under the influence of significant dark matter energy injection. We construct a novel \emph{dark main sequence}, informed, for the first time, by cutting-edge observations of the S-cluster stellar population near the Galactic Center~\cite{Ghez:2003qj, Habibi_2017}. Specifically, we discover a new branch of higher-mass dark matter powered stars that are observationally distinct from the standard main sequence. The stability of these stars is analogous to the hydrostatic equilibrium of standard stars -- as the star expands the dark matter capture rate can decrease (see App.~\ref{sec:cap}), causing the stars to contract again. The continuous accretion of dark matter fuel allows these stars to maintain equilibrium eternally, causing the dark main sequence to dominate the stellar population.

\begin{figure}[t!]
\centering
\includegraphics[width=0.49\textwidth]{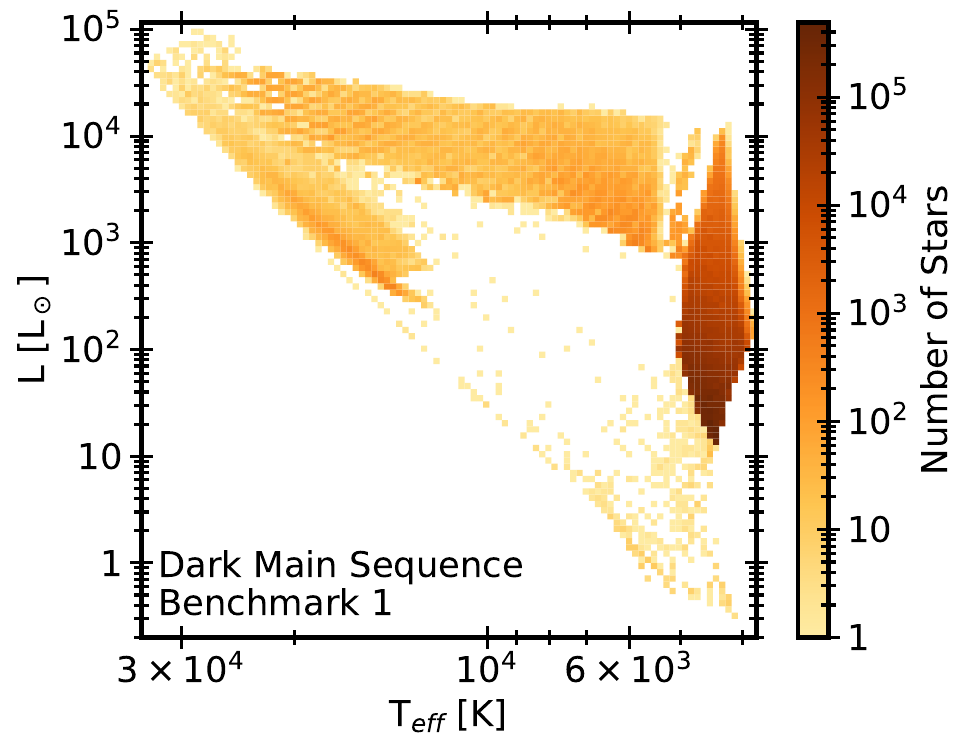}\\
\includegraphics[width=0.49\textwidth]{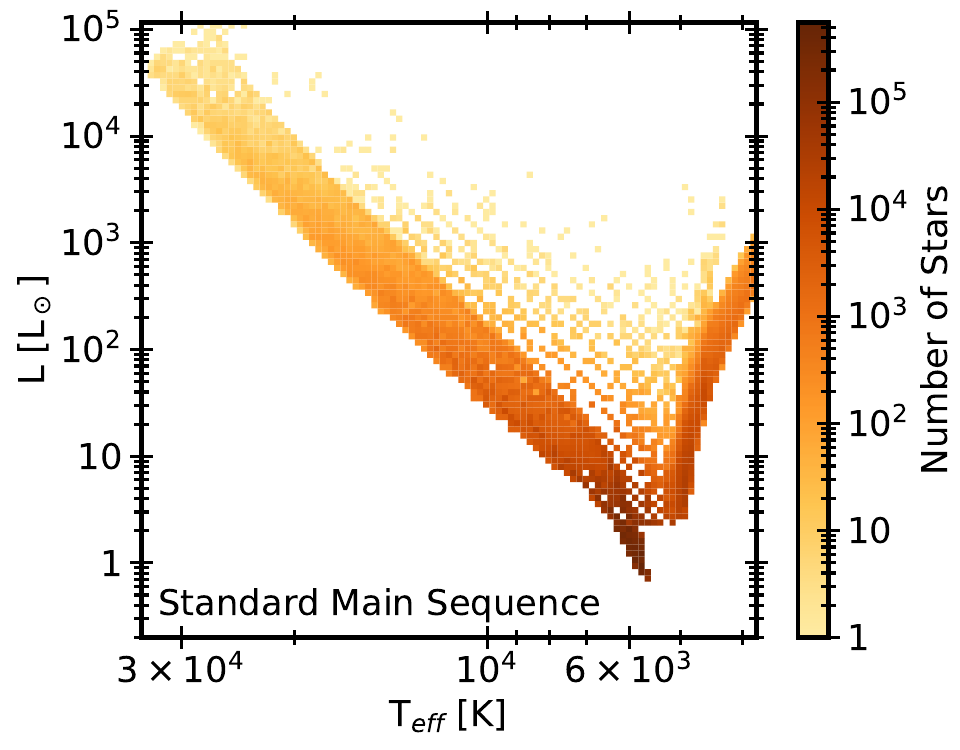}
\caption{The dark main sequence (top) compared to the standard main sequence for stellar evolution (bottom), which predicts qualitatively different stellar populations in the Galactic Center. The dark main sequence is shown for benchmark parameters of 1 GeV dark matter mass, scattering cross section of $3\times10^{-35}$~cm$^2$, and an NFW dark matter profile with $\gamma=1.5$.}
\label{fig: HR diagrams}
\end{figure}

Figure~\ref{fig: HR diagrams} shows our main result: the new dark main sequence (top), which characterizes stars very close ($<$~0.1~pc) to the Galactic Center. Compared to the standard main sequence (bottom), there are striking differences. First, dark matter powered stars can exist in stable equilibria at similar luminosities, but lower temperatures (near the Henyey track) than the standard main sequence. Second, the dark main sequence lacks low-mass stars, as they are overpowered by dark matter and move backwards along the Hayashi track~\cite{Scott:2008ns} --- potentially to the point that they are entirely disrupted~\cite{John:2023knt}.

Studying the impact of dark matter on stars at the population level allows us to statistically distinguish the dark main sequence from the standard main sequence. To date, the number of known stars in the inner parsec is limited, and more stars are needed to robustly detect any effect from dark matter. This will be possible with upcoming thirty-meter class telescopes, which will measure many new stars near the Galactic Center~\cite{TMT:2015pvw}.

Interestingly, our initial data regarding stars near the Galactic Center already unveils numerous anomalies that might be solved by the dark main sequence. Star formation models suggest that stars cannot form within $\sim$~0.1~pc of the central black hole, where the S-cluster stars are found~\cite{2012A&A...545A..70S}. Rather, the stars must have formed elsewhere and migrated towards the Galactic Center~\cite{Ghez:2003qj}. Conversely, observations suggest that stars in this region are young ($\lesssim 15$~Myr), which indicates that the stars might have formed more locally~\cite{2017ApJ...847..120H}. The tension between these two observations is known as the ``paradox of youth" problem: Galactic Center stars are as bright as young stars but show spectroscopic features of more evolved stars~\cite{Ghez:2003qj, Ghez:2003rt}. Additionally, there is a ``conundrum of old age", which describes the paucity of old and evolved stars, despite the fact that models predict the detection of many evolved stars~\cite{Buchholz_2009, Merritt_2010}. Finally, observations indicate that the initial mass function of stars near the Galactic Center is surprisingly top-heavy (biased towards high-mass stars), with observations indicating that the stellar birth distribution lies between $M^{-1.7}$~\cite{2013ApJ...764..155L} and  $M^{-0.45}$~\cite{2010ApJ...708..834B}, compared to a standard Salpeter index of  $M^{-2.35}$~\cite{1955ApJ...121..161S}. 

These anomalous observations are unique to stars at the Galactic Center, and are not well explained by conventional stellar evolution models. The very inner Galaxy is exactly where the dark matter density is high enough for dark matter annihilation to substantially replace nuclear fusion as the stellar energy source~\cite{Eisenhauer:2005cv, Scott:2008ns, Lu:2008iz, Hassani:2020zvz, John:2023knt}, allowing stars to stay forever young despite their advanced age~\cite{dylan, Scott:2008ns}. This explains the paradox of youth (because the stars have had sufficient time to migrate towards the Galactic Center)~\cite{Scott:2008ns, Fairbairn:2007bn}. We point out that this also explains the conundrum of old age (because the stars continue to reside on or near the main sequence). Because this effect is maximized in relatively high-mass stars ($\gtrsim\,$5\,$M_\odot$), while less massive stars tend to evolve backwards along the Hayashi track or be entirely disrupted, we also point out that the observed stellar distribution will naturally have a top-heavy initial mass function~\cite{2010ApJ...708..834B}. By simulating a full stellar population that includes, for the first time, the heavier stars that correspond to the recently detected stars in the innermost Galaxy, these anomalies can be explained simultaneously. \\

\section{Stellar Population Inputs}
To construct our standard and dark HR diagrams, we generate a mock stellar population that evolves both with and without the influence of dark matter. We choose relatively standard star formation parameters which are also motivated by the observation of S-cluster stars observed near the Galactic Center~\cite{2017ApJ...847..120H, 2020ApJ...899...50P}.

We start by randomly selecting stars following a Salpeter initial mass function~\cite{1955ApJ...121..161S}
\begin{equation}\label{eq: Salpeter}
\frac{dN}{dM_\star} \sim M_\star^{-2.35},
\end{equation}
where $N$ is the number of stars and $M_\star$ is the stellar mass. We consider stars with masses ranging from 1 to 20~$M_\odot$ with a step of 0.05~$M_\odot$.

We set the stellar distances to follow a radial distribution that falls as $r^{-2}$~\cite{vonFellenberg:2022lyo, 2012A&A...545A..70S, 2011ASPC..439..180H}, with 43 bins ranging from ${ 4 \times 10^{-5} }$ to ${ 4 \times 10^{-2} }$~pc. Motivated by the migration scenario, where stars cannot form closely to the inner black hole~\cite{Ghez:2003qj}, our stellar evolution models in dark matter scenarios do not include any dark matter energy injection until stars enter the main sequence, meaning that the radial binning of stars is irrelevant during their pre-main sequence life. We finally assume that star formation and migration occurs in steady state throughout the 10~Gyr lifetime of our simulations, matching the age of the Milky Way.

For simplicity, we assume that the stellar orbits are circular, $i.e.$ the dark matter capture rate is constant along the orbit. While the orbits of Galactic Center stars are highly eccentric, the equilibrium timescale for dark matter capture and annihilation is much longer than the stars' orbital period, meaning that the effective dark matter annihilation rate is only set by the time-averaged orbital radius. Thus, this only shifts the exact capture rate for each star that is found somewhere along the region around Sagittarius A* that we consider, and does not largely change our results.\\

\section{Dark Matter Inputs}
We generate two dark matter model benchmarks. Qualitatively, we find similar behavior across both, and show that similar behavior is expected for a wide range of dark matter parameters. In both cases, we assume a dark matter mass of 1~GeV, but vary the following parameters:

\paragraph*{Benchmark 1.} We set the dark matter-nucleon scattering cross section to ${ \sigma_{\chi N} = 3 \times 10^{-35} }$~cm$^2$. For this, we follow our constraints on the scattering cross section from Ref.~\cite{John:2023knt} for the star S2, which stands as the best-measured S-cluster object. To set the dark matter density along the different orbits, we choose a generalized NFW profile~\cite{Navarro:1995iw} with index ${ \gamma = 1.5 }$. This corresponds to dark matter densities of $\sim 10^{8} - 10^{12}$~GeV/cm$^3$ for the distances considered here. 

\paragraph*{Benchmark 2.} We choose a smaller dark matter-nucleon scattering cross section of ${ \sigma_{\chi N} = 1\times 10^{-37} }$~cm$^2$. For the dark matter density, we choose a standard NFW (${ \gamma = 1.0 }$), but include a density spike that arises from the adiabatic accretion of matter by Sagittarius A*~\cite{Gondolo:1999ef, Bertone:2024wbn}. We follow the spike model from~\cite{Lacroix:2018zmg, John:2023knt}, $i.e.$ ${ \gamma = 1.0 }$, ${ R_\text{spike} = 10 }$~pc and ${ \gamma_\text{spike} = 7/3 }$. This results in higher dark matter densities compared to Benchmark 1, and the density increases more steeply towards the Galactic Center, \textit{i.e.}, $\sim 10^{8} - 10^{15}$~GeV/cm$^3$.

Our benchmarks are not excluded by direct detection for spin-dependent dark matter models~\cite{XENON:2019zpr, XENON:2023cxc, CRESST:2019jnq, LZ:2022ufs, DarkSide:2018bpj}, and even given improved direct detection limits in the future, there are a wide range of additional parameter points that provide qualitatively similar results to those we show here. For example, our qualitative results are not affected for lower dark matter masses. In addition, at lower cross sections, large dark matter densities can compensate for any decrease in the signal. Finally, we note that for extremely high densities, the efficient annihilation of dark matter may cause the dark matter density to saturate at a value that depends on the dark matter mass and annihilation cross section. Even incorporating any blunting we find qualitatively new results for both $s$-wave or $p$-wave annihilation model benchmarks, see Appendix~\ref{sec: DM annihilation} for more details.\\

\section{Stellar Evolution Simulations}
We use the stellar evolution code \texttt{MESA}~\cite{2011ApJS..192....3P, 2013ApJS..208....4P, 2015ApJS..220...15P, 2018ApJS..234...34P, 2019ApJS..243...10P, 2023ApJS..265...15J}, version \texttt{r24.03.1} with Software Development Kit (SDK) version \texttt{23.7.3}, to investigate the effect of dark matter energy injection on a wide range of stellar masses. We base our model on the default \texttt{work} model, but implement a custom module to calculate the capture rate at each time step according to the current stellar radius. To obtain the capture rate, we follow the formalism of Ref.~\cite{Leane:2023woh}, under the assumption of capture-annihilation equilibrium, and add the annihilation energy into the stellar core through the \texttt{inject\_uniform\_extra\_heat} parameter. The injection of the additional energy can be restricted to a certain mass region of the star through the parameters \texttt{min\_q\_for\_uniform\_extra\_heat} and \texttt{max\_q\_for\_uniform\_extra\_heat}. We consider the stellar core to correspond to the inner 10\% of the star, but have tested that injecting the dark matter annihilation energy in the inner 1\% only results in slight numerical differences, while uniform injection throughout the star only changes our results by a factor of a few, which does not have qualitative effects. We assume that all stars initially form and evolve in a low dark matter density environment until their early main sequence phase, $i.e.$ their pre-main sequence evolution is not affected by dark matter. We assume that these stars migrate as main sequence stars towards the Galactic Center where their subsequent evolution is influenced by dark matter annihilation. As discussed above, this migration scenario is a commonly considered case, due to the fact that stellar formation close to Sagittarius A* is expected to be inhibited~\cite{Ghez:2003qj}. In \texttt{MESA}, we start injecting the additional energy when the hydrogen fraction in the star center drops below 0.67, which corresponds to an early main sequence core composition, through the parameter \texttt{xa\_central\_lower\_limit}. Once the star experiences dark matter energy injection, we evolve the star until it reaches the red giant branch phase, using the stopping condition \texttt{stop\_at\_phase\_ZACHeB}, or until an age of 10~Gyr is reached, corresponding to the age of the Galaxy. We do not include any energy transport effects directly due to internal dark matter-nucleon scattering. In our scenario these effects are subdominant, as the energy rate injected by dark matter burning far outweighs any effects of the energy transport (see \textit{e.g.}~\cite{1990ApJ...352..669G}).\\

\begin{figure*}[tbp]
\centering
\begin{minipage}[t]{0.42\textwidth}
\includegraphics[height=7cm]{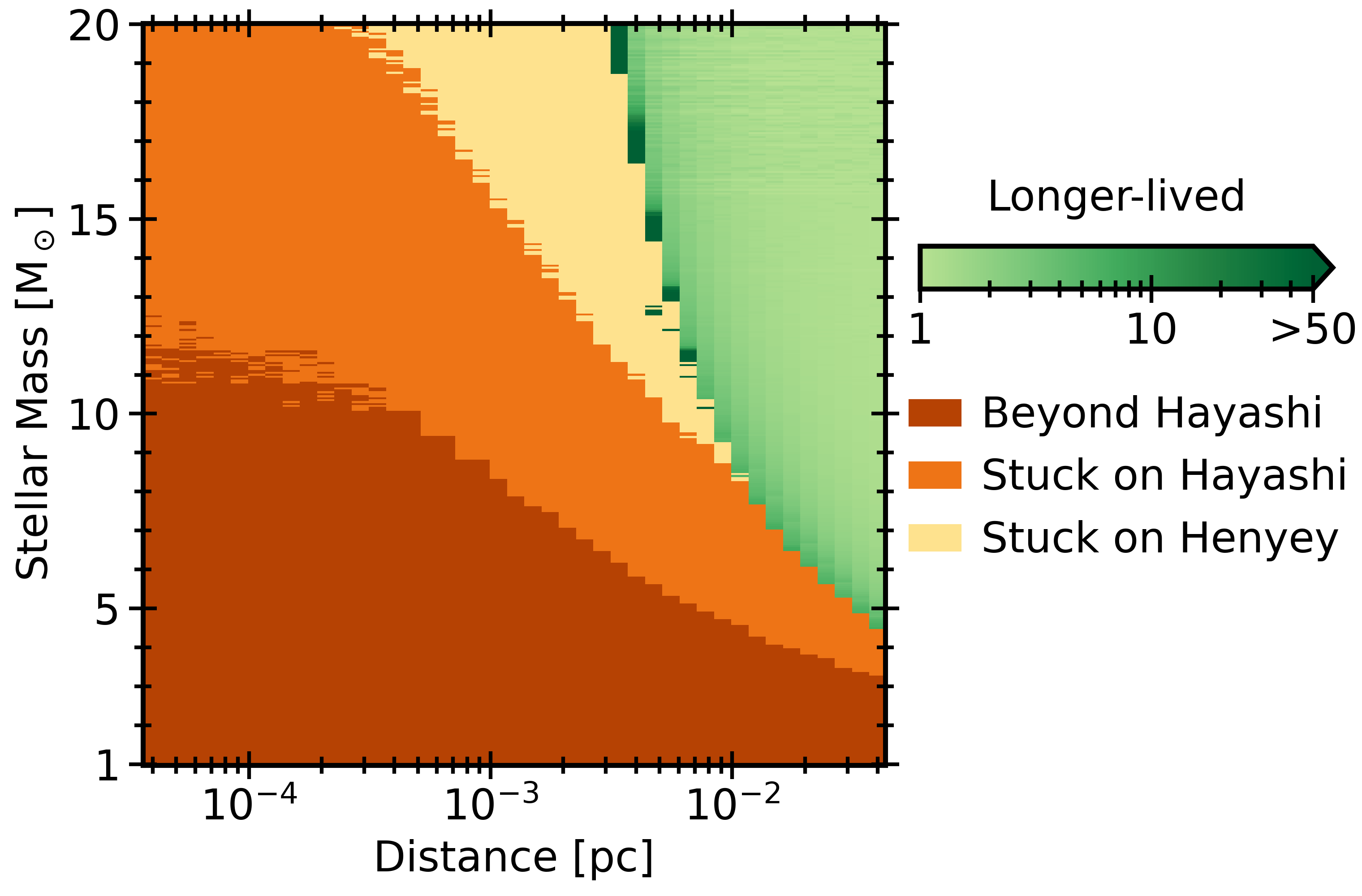}
\end{minipage}
\hfill
\begin{minipage}[t]{0.42\textwidth}
\includegraphics[height=7cm, trim={0 0 6.6cm 0},clip]{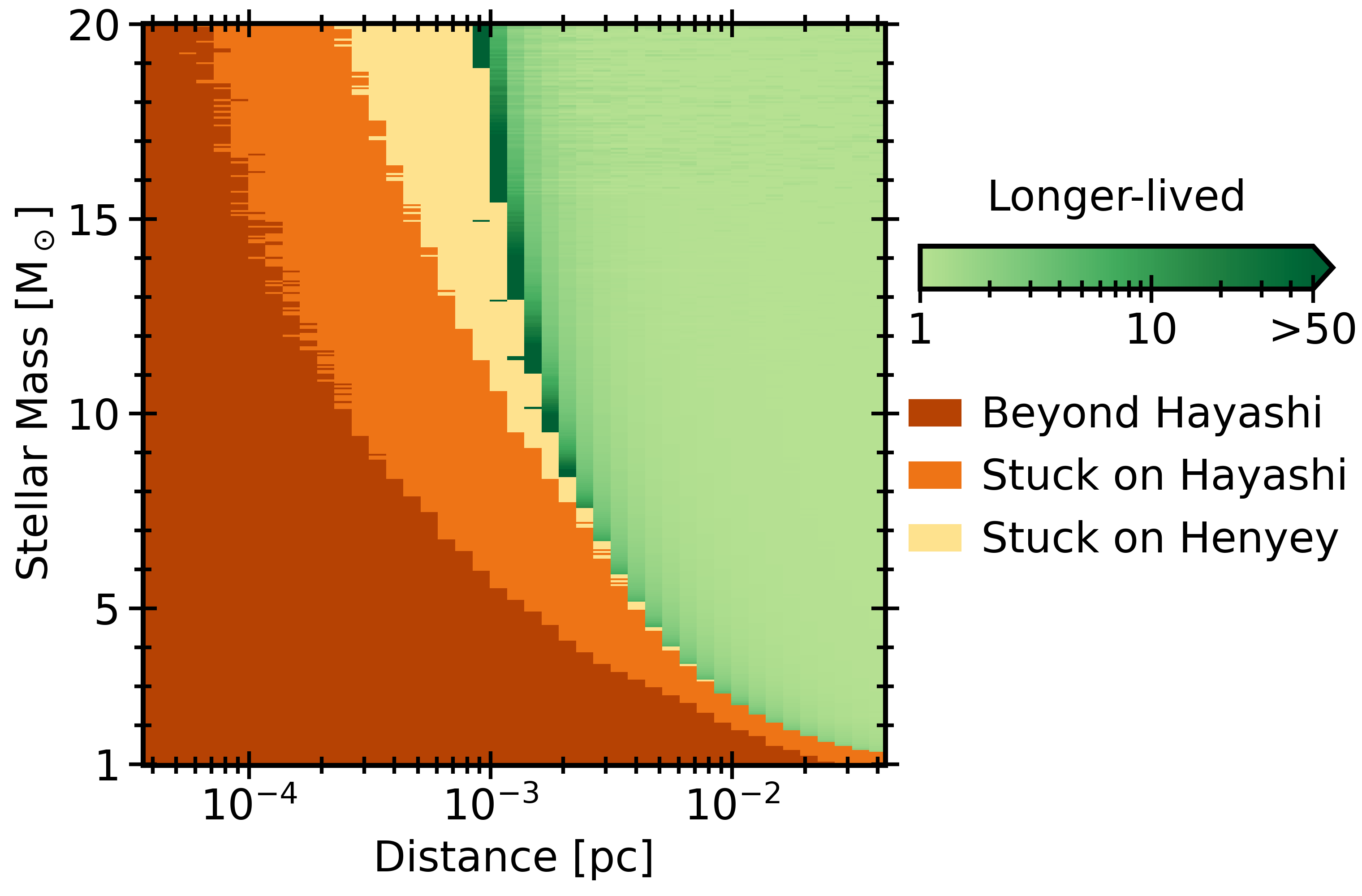}
\end{minipage}
\caption{The impact of dark matter burning on stars over a range of masses and Galactic Center distances, for Benchmark Models 1 (left panel) and 2 (right panel). Green shades indicate how much main sequence stellar evolution is slowed down by the added dark matter power compared to the standard evolution scenario. This corresponds to stars that are more massive and experience a lower dark matter density. Stars in higher density regions are significantly changed and show atypical evolution: The stars maintain equilibrium through dark matter burning and only rely on small amounts of fusion, making them reside on the Henyey track (yellow shade), or have no fusion at all, which corresponds to the Hayashi track (orange shades). This essentially ``freezes" their evolution, as stellar evolution is driven by nuclear fusion. The continuous re-supply of captured dark matter acts as an effectively infinite power source, making these stars immortal.}
\label{fig: stellar scenarios}
\end{figure*}

\section{Results and Discussions}
Figure~\ref{fig: stellar scenarios} shows the effects of dark matter energy injection on stars of different masses and distances from the Galactic Center for Benchmark 1 and 2. These models only differ in the dark matter-nucleon scattering cross section and the underlying dark matter profile, that determines the amount of dark matter a star can capture along its orbit. We find that stellar evolution is affected in various ways:

\begin{enumerate}[(1), leftmargin=17pt]

\item \textit{Longer-lived.} The green shaded regions correspond to stars that are not significantly or only moderately changed by dark matter burning. Their evolution follows the typical stages, but as part of the nuclear fusion power needed to maintain hydrostatic equilibrium is replaced by dark matter power, hydrogen in the core fuses more slowly and thus the stars' evolution is slowed down. This makes the stars longer-lived. While light green shades correspond to stars that are only slightly affected as their evolution times are similar to the standard scenario, dark green regions represent stars that can become as much as 10--100$\times$ longer-lived.

\item \textit{Stuck on Henyey track.} The yellow region corresponds to stars that are found on the Henyey track in the HR diagram, which, in the standard evolutionary scenario, is a rapid phase just before stars enter the main sequence, when hydrogen burning is ignited and increasing. This means that at the onset of dark matter burning in the stellar core, stars can move ``backwards" in evolution, returning from the main sequence to a stage with only small amounts of nuclear fusion, and maintain equilibrium primarily through dark matter burning. Dark stellar evolution on the Henyey track is extremely slow, until eventually the star does run out of hydrogen in the core and moves back to a more typical evolution on the red giant track. Thus, these stars are in principle extremely long-lived and would eventually follow the standard evolution stages, but, as in our simulations we do not evolve stars for longer than 10~Gyr, which corresponds to the Galaxy's age, these stars never move off the Henyey track. Note that many of the stars in scenario (1) that are significantly longer-lived ($\gtrsim 5$ times) also spend some time on the Henyey track, but experience a lower amount of dark matter burning and thus need to undergo more nuclear fusion, forcing them to move back to the main sequence and resume evolution earlier.

\item \textit{Stuck on Hayashi track.} The orange region corresponds to dark matter powered stars that are found in the region of the Hayashi track in the HR diagram, which, in the standard picture, describes pre-main sequence stars that are still contracting and have not yet started nuclear fusion. At the onset of dark matter burning, main sequence stars evolve ``backwards", similar to the stars halting on the Henyey track described above. In this scenario however, dark matter burning replaces nuclear fusion entirely, and stars move all the way to the Hayashi track, consistent with the results in Refs.~\cite{Fairbairn:2007bn, Scott:2008ns}. The stars do not undergo any further evolutionary stages as long as the dark matter capture rate remains sufficiently high to prevent nuclear fusion from resuming. A star like this might look like a young, still-forming star but has features of a star that has undergone nuclear fusion in the past -- and is effectively immortal.

\item \textit{Beyond Hayashi track.} The dark orange region corresponds to stars moving back onto the Hayashi track, similar to case~(3). However, in this case, the stars move further up the Hayashi track to even earlier evolutionary states. In our \texttt{MESA} simulations, these stars end up higher on the Hayashi track than their starting position. We emphasize that this scenario is highly reliant on the modelling methods (as \texttt{MESA} does not simulate the very early stages of star formation). It is possible that these stars could expand even further or be entirely disrupted by the dark matter energy injection. A scenario like this may be consistent with G-objects, a recently discovered strange class of objects found at similar orbits around Sgr A* as the S-cluster stars that resemble clouds of gas surrounding a denser core~\cite{2017ApJ...847...80W, 2020Natur.577..337C}.
\end{enumerate}

In both benchmarks, massive stars further out from the Galactic Center are the least affected, while light stars can become immortal even at distances where the dark matter density is not sufficient to significantly affect more massive stars. The different dark matter density profiles are also reflected in Fig.~\ref{fig: stellar scenarios}: While Benchmark 1 (left panel) uses an NFW profile, Benchmark 2 (right panel) includes a dark matter spike that steepens the profile and, accordingly, the region of the parameter space in which stars are immortal.

\begin{figure}[t!]
\includegraphics[width=\columnwidth]{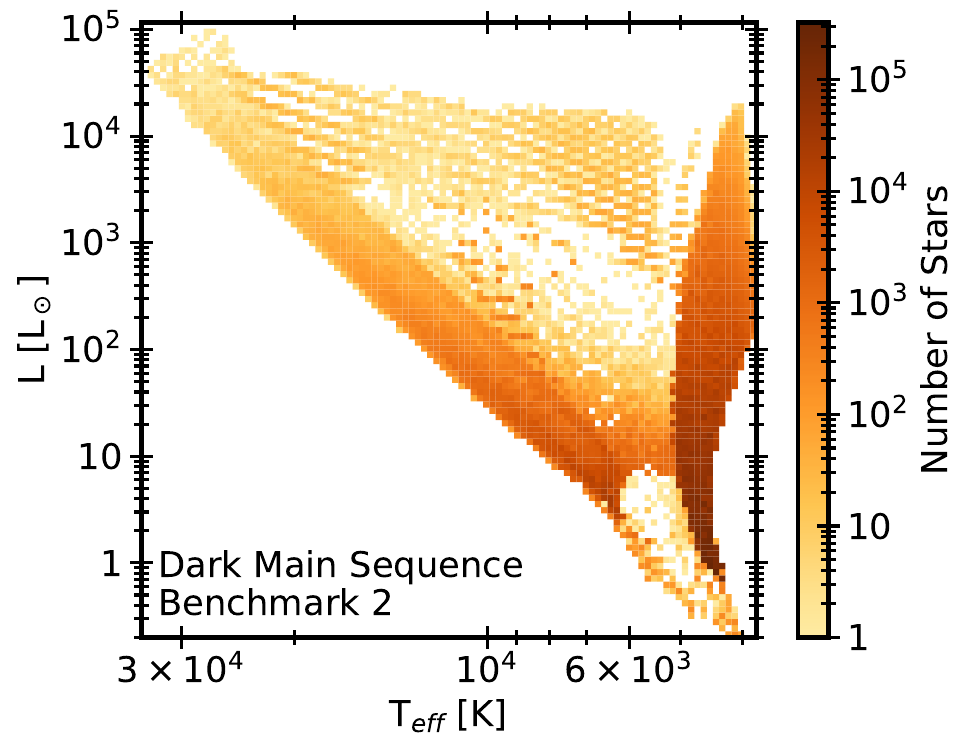}
\caption{Dark HR diagram for a population of Galactic Center stars, for our dark matter benchmark 2, with 1 GeV dark matter mass, scattering cross section of $10^{-37}$~cm$^2$, and NFW dark matter profile with a spike (see text for details).}
\label{fig: HR diagram spike}
\end{figure}

Figure~\ref{fig: HR diagrams} shows HR diagrams for a population of Galactic Center stars with and without dark matter for Benchmark~1, and Figure~\ref{fig: HR diagram spike} shows the dark HR diagram for the additional Benchmark~2. We include evolutionary stages from the early main sequence up until the stars leave the main sequence and start hydrogen shell burning. We do not include stars that have become red giants and start core Helium burning, $i.e.$ most of the stars on the HR diagram are main sequence stars. For more details, see Appendix~\ref{sec: computation of HR diagrams}. Most stars lie on the main sequence, $i.e.$ the diagonal line, where more massive stars are at the top left, as they are hotter and more luminous, and lighter stars are at the bottom right. The small number of stars scattered to the right of the main sequence are stars that are older stars that have just moved away from the main sequence towards the red giant branch; which manifests in a lower temperature compared to the main sequence. 

For our dark main sequence in Figs.~\ref{fig: HR diagrams} and~\ref{fig: HR diagram spike}, the effect of dark matter burning results in strikingly distinct features. Compared to the standard main sequence, the Hayashi tracks (approximately vertical tracks at low temperatures) are densely populated with immortal stars (see scenarios (3) and (4) above), and the Henyey tracks (approximately horizontal tracks at high luminosities), are clearly visible (see scenario (2) above), containing stars that are mostly powered by dark matter burning with only small amounts of nuclear fusion, making their evolution extremely slow. Additionally, the region where light main sequence stars are located (lower luminosities and temperatures), are only sparsely populated, as most of these stars have moved to the Hayashi tracks. In Benchmark~2 (Fig.~\ref{fig: HR diagram spike}), more of these low-mass stars remain on the main sequence compared to Benchmark 1 (Fig.~\ref{fig: HR diagrams}), due to the steeply increasing dark matter density of the spike model, as can also be seen in Fig.~\ref{fig: stellar scenarios}. We note that the striped features observed in the Henyey branch are due to our distance (and thus dark matter density) binning, which sets the stellar luminosity, and are thus not a physical feature. 

Finally, we note that the fraction of high-mass stars increases in dark matter scenarios due to the effect of stellar immortality. If only standard stellar models were considered, the large fraction of high-mass stars would be misconstrued as evidence of a top-heavy initial mass function (IMF). Intriguingly, observations have found evidence of a top-heavy IMF at the Galactic Center~\cite{2013ApJ...764..154D, 2013ApJ...764..155L, 2010ApJ...708..834B}. We find that dark matter would harden the IMF, compared to a Salpeter IMF, by approximately ${ \sim 0.5 - 2.5 }$, as shown in Figs.~\ref{fig: HR diagrams} and~\ref{fig: HR diagram spike}.\\

\section{Summary and Outlook}
We have produced the first population-level prediction for stellar evolution under the presence of dark matter capture and annihilation near the Galactic Center. We have established a new ``dark main sequence" that is observable in the Hertzsprung-Russell diagram of stars observed in dark matter rich environments. This sequence has three observable characteristics that clearly separate it from standard main sequence stellar evolution: (1) a top-heavy initial mass function along the main sequence branch, produced by the efficient removal of low-mass stars off of the main sequence and into locations on the Hayashi or Beyond-Hayashi tracks, (2) a bright and densely populated Hayashi track, produced by low-mass main sequence stars which achieve eternally stable equilibria in configurations with large stellar radii and low stellar temperatures, (3) a bright and densely populated Henyey track at higher-masses, populated by effectively immortal (typical lifespans $>$10~Gyr) high-mass stars with similar luminosities but slightly lower temperatures than their main sequence configurations. 

Excitingly, there are several anomalies in Galactic Center stellar observations which can be simultaneously explained by stellar evolution along our dark main sequence. The top-heavy initial mass function near the Galactic Center is automatically produced by our simulations with dark matter, while both the paradox of youth (see also~\cite{Scott:2008ns}) and conundrum of old age can be explained if relatively high-mass stars are powered indefinitely by dark matter annihilation. Finally, observations of G-objects (extremely large gas clouds which may hold a central star), may be consistent with Beyond-Hayashi evolution, although the stability of such solutions falls beyond the parameters of our current \texttt{MESA} modeling. We will investigate this in upcoming work. 

Going forward, our stellar population predictions have important implications for the extremely difficult observations of Galactic Center stars. The vastly cooler temperatures of Hayashi track stars may impede their detection unless observational efforts specifically look for such objects. Additionally, the dark Henyey track stars may be mischaracterized as standard main sequence stars if the possibility of Henyey track evolution is not considered. Verifying or ruling out dark main sequence evolution will require the detection of new stars for sufficient sampling and to sufficiently populate the dark main sequence. This will be possible with new telescopes, such as thirty-meter class telescopes, or with additional stars measured with the Very Large Telescope (VLT) or the Keck Observatory. \\

\section*{Acknowledgements}
We thank Andrea Ghez for helpful discussions. This work was supported by a collaborative visit funded and hosted by Stanford KIPAC and the Cosmology and Astroparticle Student and Postdoc Exchange Network (CASPEN). I.J. and T.L. acknowledge support by the Swedish Research Council under contract 2022-04283. T.L. also acknowledges support from the Swedish National Space Agency under contract 117/19. R.K.L. is supported by the U.S. Department of Energy under Contract DE-AC02-76SF00515. This project used computing resources from the National Academic Infrastructure for Supercomputing in Sweden (NAISS) under project NAISS 2023/3-21.\\

\appendix
\section{Dark Matter Capture Rates for Actively Evolving Stars}
\label{sec:cap}

\begin{figure*}[tbp]
\centering
\includegraphics[width=0.49\textwidth]{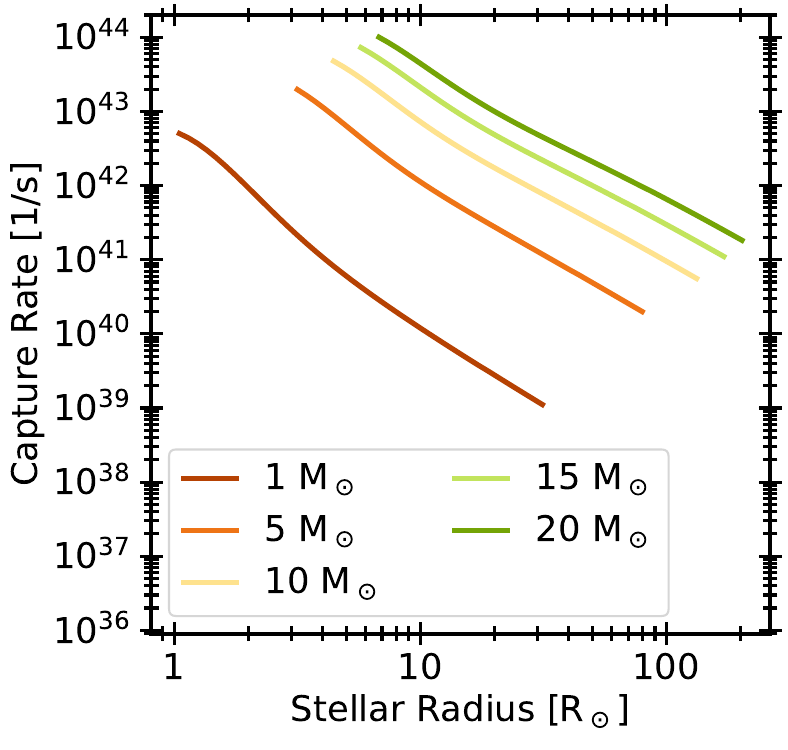}
\includegraphics[width=0.49\textwidth]{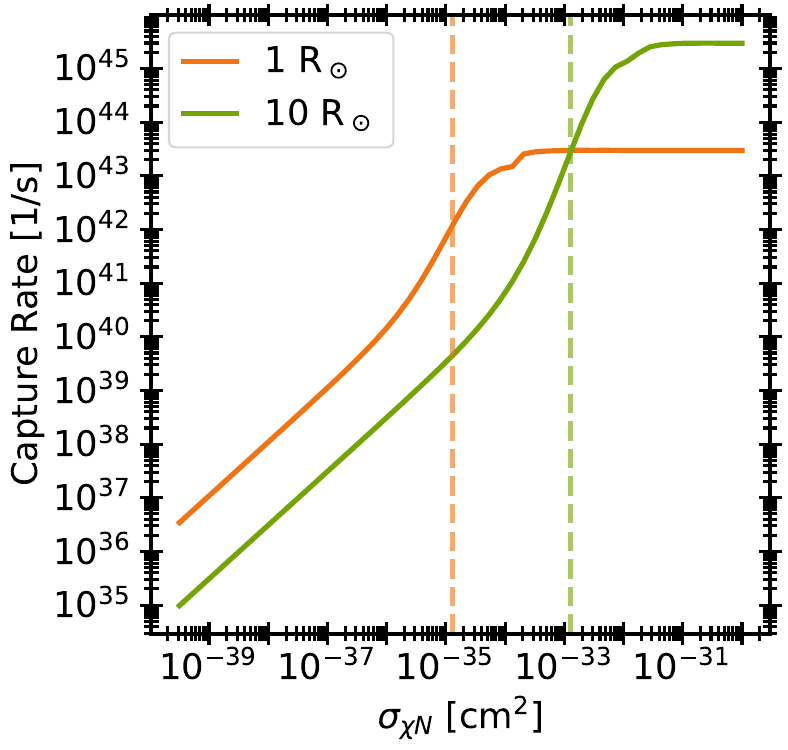}
\caption{Left panel: Dark matter capture rate as a function of stellar radius, for a scattering cross section of $3\times 10^{-35}$~cm$^2$ (Benchmark 1). Right panel: Capture rate as a function of scattering cross section, for stellar radii of 1 and 10 solar radii, and a fixed stellar mass of 1 solar mass. The dashed line shows the transition cross sections for the two stars. To calculate these capture rates, we assume an example dark matter mass of 1~GeV, orbital velocity of $10^9$~cm/s and dark matter density $2\times10^{12}$~GeV/cm$^3$.} 
\label{fig: capture rate vs stellar radius}
\end{figure*}

Figure~\ref{fig: capture rate vs stellar radius} (left panel) shows the dark matter capture rates expected as the star undergoes evolution, $i.e.$ taking into account the changing mass and radius of the star. The first point to note, as highlighted by the left figure, is that the capture rate changes as the stellar radius changes, and so it is important to take the updated stellar radius into account for the capture rate as the star evolves. This implementation is an improvement over previous studies~\cite{John:2023knt}. The second point is that for a fixed stellar mass and fixed dark matter scattering cross section, the dark matter capture rate can \textit{decrease} as the stellar radius increases. The reason for this can be understood from the right figure.

Figure~\ref{fig: capture rate vs stellar radius} (right panel) shows the capture rate as a function of scattering cross section, for two example stars of radius 1 and 10 times the solar radius, for a fixed stellar mass of 1 solar mass. We also show as the dashed line the transition cross section for these two example stars. 
Here, the transition cross section is $\sigma_{\rm tr}=\pi R^2/N$, where $N$ is the number of stellar nucleons available for scattering. Physically, this corresponds to a mean free path of the dark matter of about the size of the star, and therefore corresponds to the cross section above which multiple scatters can be needed for capture to occur. Below this cross section, dark matter can only be captured with roughly one scatter, as the scattering is so weak that on average one or zero scatters occur. As the mass of these two example stars stays the same, the number of targets also remains the same, except they are distributed over a larger volume, the transition cross section between the two simply scales as their radius ratio squared.

As Fig.~\ref{fig: capture rate vs stellar radius} (right) shows, larger radii stars do capture more dark matter in the sufficiently-high cross section regime where roughly all incoming dark matter is captured, as their surface areas available for capture are larger. However, less dense objects have lower escape velocities, so require more scatters (i.e. a larger cross section) to bring the dark matter to a low enough speed to be captured, which is why the larger star needs a larger cross section increase above its transition cross section at $\sim 10^{-33}$~cm$^2$ to reach its geometric capture rate. In comparison, the smaller radius star increases less above its $\sim 10^{-35}$~cm$^2$ transition before reaching geometric capture, because it is more efficient at capture (and so needs fewer scatters) due to its higher escape velocity. Once comparing the capture rates for these two objects at their transition cross section, the puffier star is already down a larger penalty compared to its geometric maximum capture than the more dense star, which is then cancelled out by its size, such that at their respective transition cross sections (which are two orders of magnitude apart) they actually have the same capture rate. This means that, because the capture rate decreases linearly with cross section below the transition cross section, the puffier, larger radius star has a lower capture rate at smaller cross sections, as its transition cross section is higher to start with.
This can lead to interesting effects during the stellar evolution process as the star becomes larger, as the dark matter capture and annihilation driving the growth in radii consequently decreases.

For our benchmark, we assume all dark matter remains captured, $i.e.$ that it does not evaporate. This is a good approximation for the stars we consider, in the dense Galactic Center environment.

\section{Dark Matter Annihilation Rates}
\label{sec: DM annihilation}

\subsection{Impact of $s$-wave vs $p$-wave Annihilation on the Dark Matter Density}

Depending on the size of the dark matter annihilation rate, the dark matter density can saturate, \textit{i.e.} not increase further, towards the Galactic Center. In the main text we do not show the impacts of this blunting, for completeness in this section we will now investigate results for two common scenarios, $s$-wave or $p$-wave annihilation, both of which can arise in a broad range of particle models. We will show that the variance in the expected dark-matter density provides an interesting handle on the size of the dark matter annihilation rate, through different stellar evolution behavior on our dark HR diagrams, with striking qualitatively different behavior compared to the standard HR diagram in all cases.

This saturation density due to dark matter annihilation in the Galaxy is given by
\begin{equation}\label{eq: saturation density}
\rho_\text{sat} = \frac{m_\text{DM}}{\langle\sigma v\rangle T_\text{BH}},
\end{equation}
where $m_\text{DM}$ is the mass of the dark matter particle, $\langle\sigma v \rangle$ is the dark matter annihilation cross section, and ${ T_\text{BH} \sim 10^{10} }$~yr is the age of Sgr A*, which corresponds to the time dark matter has been accumulated at the Galactic Center. 

\subsubsection{S-wave Annihilation}
\begin{figure*}[t]
\centering
\includegraphics[width=0.49\textwidth]{HR_standard.pdf}
\includegraphics[width=0.49\textwidth]{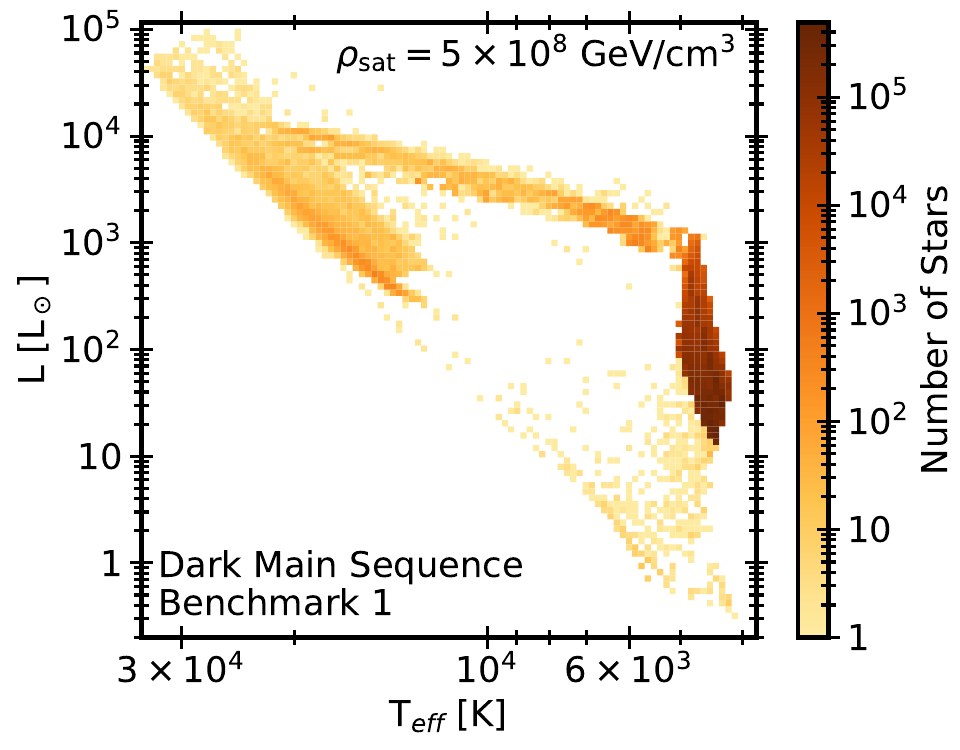}
\caption{Similar to Fig.~\ref{fig: HR diagrams}, where the left panel shows the standard main sequence for stars without dark matter burning, while the right panel presents the a case similar to our Benchmark 1 scenario, but with a dark matter density that saturates at $\sim 5 \times 10^8$~GeV/cm$^3$ towards the Galactic Center, which is applicable to model classes with $s$-wave annihilation. Compared to the un-saturated case in Fig.~\ref{fig: HR diagrams}, the dark main sequence is reduced but still clearly visible, and regions of the lighter stars show similar differences as compared to the standard HR diagram.}
\label{fig: HR diagram benchmark 1 DM saturation}
\end{figure*}

Figure~\ref{fig: HR diagram benchmark 1 DM saturation} illustrates an example of the stellar evolution for classes of particle models with $s$-wave annihilation rates ($\langle\sigma v\rangle = 3\times10^{-26}$~cm$^3$/s), at an example mass of 1 GeV, corresponding to a dark matter saturation limit of $\sim 10^8$~GeV/cm$^3$. For comparison, the left panel shows the standard HR diagram, while the right panel shows the dark HR diagram. We observe that the dark HR diagram shows striking differences compared to the standard HR diagram. Intermediate mass stars ($\sim 8 - 13$~$M_\odot$), are observed to populate a distinct dark main sequence (formed from stars residing along the Henyey tracks), albeit thinner than in the non-saturated case (see Fig.~\ref{fig: HR diagrams}). Even lighter stars are disrupted already at much lower dark matter densities of $\sim 10^6$~GeV/cm$^3$, which is already achieved for the standard NFW profile ($\gamma = 1$)~\cite{John:2023knt}. This means that even at much lower densities than considered in our Benchmark 1 and 2 scenarios (Figs.~\ref{fig: HR diagrams} and~\ref{fig: HR diagram spike}), the dark HR diagram displays clear differences to the standard scenario, lacking light stars on the main sequence and strongly populating the Hayashi tracks.

\subsubsection{P-wave Annihilation}

In the case of particle model classes with $p$-wave annihilation ($\langle\sigma v\rangle \approx 10^{-31}$~cm$^3$/s), the saturation density in the Galaxy is much higher than $s$-wave cases, due to the lower annihilation rates in the Galaxy (see Eq.~\ref{eq: saturation density}). Assuming for example an NFW dark matter profile~\cite{Navarro:1995iw} with a profile index of $\gamma = 1.5$, the maximum dark matter density, achieved at the innermost location of S-cluster stars, corresponds to $\sim 10^{12}$~GeV/cm$^3$. This would produce the same dark HR diagrams as presented in the main text, as the blunting is then not important at the Galactic radii of the S-cluster stars.  While we observe qualitatively different behaviour for both the $s$-wave and $p$-wave results compared to the standard HR diagram, comparing Fig.~\ref{fig: HR diagrams} with Fig.~\ref{fig: HR diagram benchmark 1 DM saturation}, it is also clear that our setup can provide powerful insights into the dark matter density distribution and dark matter self-annihilation rate, through a statistically different number of stars being affected the same way.

\subsection{Dark Matter Equilibrium}

Throughout our analysis, we assume that the dark matter capture rate and annihilation rate inside the star are in equilibrium. As we assume that stars form in a region of low dark matter density and migrate towards the Galactic Center during their main sequence phase, some time passes until the dark matter capture rate and annihilation rate inside the star reach equilibrium. This equilibrium time scale is estimated by~\cite{Bramante:2023djs}
\begin{equation}\label{eq: equilibrium time scale}
\tau_\text{eq} = \sqrt{\frac{V_\star}{C \langle\sigma v\rangle}},
\end{equation}
where $V_\star = 4/3 \times \pi R_\star^3$ is the volume of the star for a stellar radius $R_\star$ and $C$ the dark matter capture rate~\cite{Leane:2023woh, John:2023knt}. Thus, the equilibrium time scale depends on the annihilation cross section (as well as on the dark matter density and other parameters through the capture rate).

In the most extreme case we can consider the above derived required annihilation cross section of $\sim 10^{-31}$~cm$^3$/s, for which the equilibrium time scale is $\sim 10^6$~yr. The age of the S-cluster stars is not exactly known, but estimated to be in the order of a few to a few hundreds of~Myr~\cite{Habibi_2017} -- however, if dark matter extends the lifetime of the stars, they could be much older. Therefore, we expect our assumption of equilibrium to be valid.

\section{Computation of Hertzprung-Russell Diagrams}
\label{sec: computation of HR diagrams}
We now provide our methodology to obtain the HR diagrams shown in Figs.~\ref{fig: HR diagrams},~\ref{fig: HR diagram spike} and~\ref{fig: HR diagram benchmark 1 DM saturation} from our simulated stellar population, as shown in Fig.~\ref{fig: stellar scenarios}.

In our HR diagrams, we only include main sequence stars, \textit{i.e.} we do not show evolutionary stages before and after the main sequence (as we defined in \texttt{MESA}). This is because the S-cluster stars are classified as main sequence stars and we want to re-create HR diagrams that would closely resemble the S cluster.

We then populate the HR diagram by selecting stars according to (1) an initial mass distribution following the Salpeter distribution in Eq.~\ref{eq: Salpeter}, (2) a radial distribution of distances from the Galactic Center (corresponding to different dark matter densities, which affects the stars differently) and (3) an age from 0 to 10~billion years. If for a star of a selected age the current evolutionary stage does not correspond to the main sequence, we do not include this star on the HR diagram, \textit{i.e.} we do not include the star if it has evolved beyond the main sequence or if it is too young to have reached the main sequence. 

Note that when obtaining the HR diagram for the standard main sequence (Fig.~\ref{fig: HR diagrams} bottom panel), the radial distance does not matter, since in the standard scenario we assume that stellar evolution is not changed by dark matter.

\bibliography{main}

\end{document}